\documentclass[journal]{IEEEtran}
\usepackage{amsmath,amsthm}
\usepackage{cite}
\usepackage{graphicx}
\usepackage{epstopdf}
\usepackage{color}
\usepackage{amsfonts,amsmath,amssymb}
\usepackage{cite}
\usepackage{graphicx}
\usepackage{url}
\usepackage{bm}
\usepackage{bbm}
\usepackage{amssymb}
\usepackage[T1]{fontenc}
\usepackage{fancyhdr}
\begin{document}

\title{Photonic Engineering for CV-QKD over Earth-Satellite Channels}
\author{
\IEEEauthorblockN{Mingjian He$^1$,  Robert Malaney$^1$, and Jonathan Green$^2$}\\

\IEEEauthorblockA{$^1$School of Electrical Engineering  \& Telecommunications,\\
The University of New South Wales,
Sydney, NSW 2052, Australia. \\
$^2$Northrop Grumman Corporation, San Diego,
California, USA. }}


\maketitle
\thispagestyle{fancy}
\renewcommand{\headrulewidth}{0pt}


\begin{abstract}
Quantum Key Distribution (QKD) via satellite offers up the possibility of unconditionally secure communications on a global scale. Increasing the secret key rate in such systems, via photonic engineering at the source, is a topic of much ongoing research.
In this work we investigate the use of  photon-added states and photon-subtracted states, derived from two mode squeezed vacuum states, as examples of such photonic engineering. Specifically, we determine which engineered-photonic state provides for better QKD performance when implemented over channels connecting terrestrial receivers with Low-Earth-Orbit satellites. We quantify the impact the number of photons that are added or subtracted has, and highlight the role played by the adopted model  for atmospheric turbulence and loss on the predicted key rates.
Our results are presented in terms of the complexity of deployment used, with the simplest deployments ignoring any estimate of the channel, and the more sophisticated deployments involving a feedback loop that is used to optimize the key rate for each channel estimation. The optimal quantum state  is identified for each deployment scenario investigated.
\end{abstract}

\section{Introduction}
Quantum Key Distribution (QKD) via satellite offers a paradigm shift for the deployment of large-scale quantum information protocols, e.g. \cite{bedington2017progress,hosseinidehaj2015gaussian,hosseinidehaj2015quantum}. Light propagation through the atmosphere to and from Low-Earth-Orbit (LEO) satellites can overcome the $\sim 100$km distance limitation  that currently constrains terrestrial quantum communication  links. In the past few years  breakthroughs have been made in the actual deployment of quantum communications via satellites \cite{liao2017satellite,yin2017satellite,ren2017ground,takenaka2017satellite} - breakthroughs that use Discrete-Variable (DV) technology (single-photon states and entangled photon pairs).

Continuous-Variable (CV) quantum states provide an entirely different way to transmit  quantum information. Compared to DV technology,  CV technology has the advantage of `off-the-shelf' technology based on efficient  homodyne detectors \cite{grosshans2003quantum,scarani2009security} (or polarization detectors \cite{vidiella2006continuous}), and the fact that QKD protocols using CV detectors potentially provides for a more realistic route to higher secret key rates \cite{pirandola2015mdi}.

Currently, no experimental implementation of space-based CV-QKD has been carried out. But this is expected to change in the near future\cite{hosseinidehaj2018satellite}. Being a special category of CV quantum states,  Gaussian CV states have been well-researched both theoretically and experimentally \cite{weedbrook2012gaussian}. However, CV-QKD protocols with  non-Gaussian states, such as those produced via photon subtraction, have also garnered great interest. This is so partially because  such states potentially allow for a higher level of entanglement at a given energy \cite{navarrete2012enhancing,kitagawa2006entanglement,zhang2010distillation}, and that they can be a  pivotal resource for quantum information tasks such as quantum error correction \cite{Nogo}. It is natural to hypothesize that  non-Gaussian states can also boost the secret key rate of CV-QKD protocols, e.g.,
\cite{huang2013performance,zhao2017improvement,lim2017continuous, ma2018continuous}.

In our previous work we have investigated the performance of a CV-QKD protocol using a single-photon-subtracted state over an Earth-satellite channel, determining whether transmitter or receiver photonic subtraction is preferred \cite{he2018quantum}. In this work we significantly extend our previous study with the following new contributions.  (i) Multiple-photon-subtracted states are investigated. These highly non-Gaussian states are created by performing multiple-photon subtraction to a Two Mode Squeezed Vacuum (TMSV) state at the transmitter. (ii) Multiple-photon-added states are also investigated, again via photonic engineering on the TMSV state at the transmitter. (iii) Optimization of the  input TMSV states (dynamic adjustments to squeezing) based on the anticipated loss in the channel is investigated. (iv) An updated model for the probability density function (PDF) of the transmissivity of the fading channel is adopted \cite{vasylyev2016atmospheric}, which includes other important physical effects beyond the beam-wandering effect used in  our previous work.

 Looking at all these effects collectively will allow us to determine  whether the use of photonic engineering at the transmitter truly allows for better pathways to improved higher secret key rates over Earth-satellite channels. No previous study has looked at all the above effects collectively in the manner we do here.

The structure of the remainder of this paper is as follows. In Section \ref{channelmodel}, the model of the Earth-satellite quantum channel between terrestrial stations and LEO satellites is described. In Section \ref{PSSandPAS}, our CV-QKD protocols with non-Gaussian states are described, and in Section \ref{results} of our simulation results are presented.

\vskip 20pt
\section{Earth-satellite Channels}\label{channelmodel}
\begin{figure}
	\includegraphics[width=.48\textwidth]{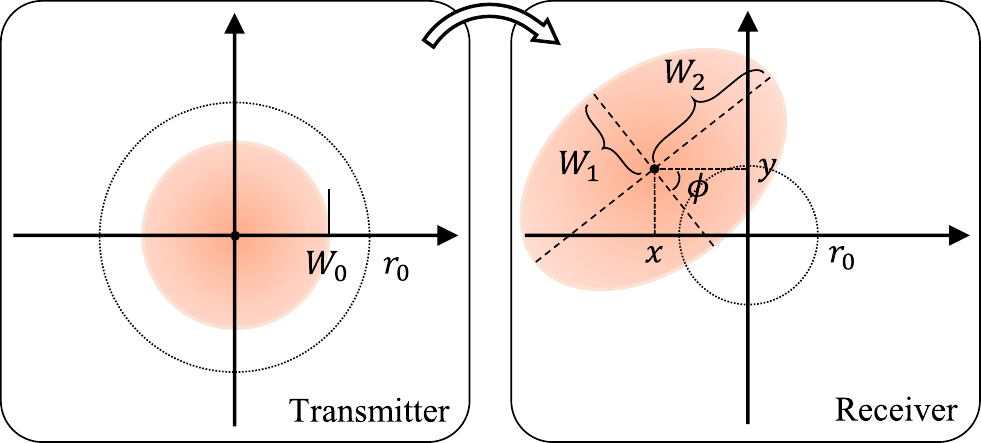}
	\centering
	\caption{Evolution of the beam-profile over the Earth-satellite channel, where beam-wandering causes the deviation of the beam-centroid, beam-broadening expands  the beam-profile, and beam-deformation alters the shape of the beam-profile. Here the orange circle represents the beam-profile at the transmitter, the dotted circle with radius $r_0$ illustrates the detector aperture, and the orange eclipse represents the beam-profile at the receiver. Here we use different scales for the two sides for better illustration.}
	\label{beam}
\end{figure}

We consider the model of a direct vertical link between the  satellite and a terrestrial station. Our quantum information carrier is a pulsed optical beam. For optical signals in the Earth-satellite channel, the dominant loss mechanisms will be  beam-wandering, beam-broadening, and  beam-deformation, all randomly caused by turbulence in the Earth's atmosphere (Fig.~\ref{beam}) \cite{andrews2005laser}. The beam-broadening is also a consequence of diffraction.
These effects are well-described by a recent model proposed by \cite{vasylyev2016atmospheric}, where the channel transmissivity (aperture transmittance) $T_E$ reads
\begin{equation}
T_E = T_0\exp\left\lbrace-\left[\frac{\sqrt{x^2+y^2}/r_0}{R(\frac{2}{W_{\rm{eff}}(\phi-\phi_0)})}\right]^{ \lambda\left(2/W_{\rm{eff}}(\phi-\phi_0)\right) }\right\rbrace\textrm{,}
\label{Cheq1}
\end{equation}
where $(x,y)$ is the 2-D position of the beam-centroid, $r_0$ is the aperture radius of the detector, $\phi$ is the beam rotation angle, $\phi_0=\tan^{-1}\frac{y}{x}$,  $W_{\rm{eff}}$ is the effective spot-radius, and $T_0$ is the maximal attainable transmissivity achieved when $(x,y)=(0,0)$ (i.e. no beam-centroid deviation). These latter two parameters can be expressed
\begin{equation}
\begin{array}{*{20}{l}}	
W_{\rm{eff}}^2(\phi)= & 4r_0^2\left\lbrace\mathcal{W}\left(\frac{4r_0^2}{W_1W_2}e^{(r_0^2/W_1^2)\left[1+2\cos^2(\phi)\right]}\right.\right.\\ &
\times\left.\left. e^{(r_0^2/W_2^2)\left[1+2\sin^2(\phi)\right]}\right)\right\rbrace^{-1}\textrm{,}
\end{array}
\end{equation}
\begin{equation}
\begin{array}{*{30}{l}}
T_0 =
& 1 - I_0\left(r_0^2\left[\frac{1}{W_1^2}-\frac{1}{W_2^2}\right]\right)e^{-r_0^2\left(1/W_1^2+1/W_2^2\right)}\\
& - 2\left[1-e^{-(r_0^2/2)\left[1/W_1-1/W_2\right]^2}\right] \\
&\times\exp\left\lbrace-\left[\frac{\frac{(W_1+W_2)^2}{\left|W_1^2-W_2^2\right|}}{R\left(\frac{1}{W_1}-\frac{1}{W_2}\right)}\right]^{\lambda\left(\frac{1}{W_1}-\frac{1}{W_2}\right)}\right\rbrace\textrm{,}
\end{array}
\end{equation}
where  $W_1$ and $W_2$ are elliptical semi-axis lengths, and $R(W)$ and $\lambda(W)$ are scaling and shaping functions given by,
\begin{equation}		
R(W) = \left[\ln\left(2\frac{1-\exp\left[-\frac{1}{2}r_0^2W^2\right]}{1-\exp\left[-r_0^2W^2\right]I_0(r_0^2W^2)}\right)\right]^{-\frac{1}{\lambda(W)}}\textrm{,}
\end{equation}	

\begin{equation}
\begin{array}{*{20}{l}}
\lambda(W) = & 2r_0^2W^2\frac{\exp{[-r_0^2W^2]}I_1(r_0^2W^2)}{1-\exp\left[-r_0^2W^2\right]I_0(r_0^2W^2)} \\ &
\times
\left[\ln\left(2\frac{1-\exp\left[-\frac{1}{2}r_0^2W^2\right]}{1-\exp\left[-r_0^2W^2\right]I_0(r_0^2W^2)}\right)\right]^{-1}\textrm.
\end{array}
\label{Cheq5}
\end{equation}
Here $\mathcal{W}(\cdot)$ is the Lambert W function, and $I_i(\cdot)$ is the modified Bessel function of $i$-th order.


The channel transmissivity $T_E$ is a function of five real random variables $\left\lbrace x,y,\theta_1,\theta_2,\phi\right\rbrace$, where $\theta_i=\ln\frac{W_i^2}{W_0^2}$ with $W_0$ being the beam-waist. It is assumed that $x$ and $y$ are i.i.d. and they both follow a zero-mean Gaussian distribution. Parameters $\theta_1$ and $\theta_2$ are taken to follow a joint-Gaussian distribution. Assuming the turbulence is isotropic, the rotation angle $\phi$ is uniformly distributed.

Building upon \cite{vasylyev2016atmospheric}, \cite{guo2018channel} showed for Earth-satellite links the mean and variance of $\left\lbrace x,y,\theta_1,\theta_2\right\rbrace$ can be written
\begin{gather}
\begin{aligned}
&\left< \theta_{1/2} \right>= \ln{\left[\frac{\left(1+2.96\sigma_I^2\Omega^{5/6}\right)^2}{\Omega^2\sqrt{\left(1+2.96\sigma_I^2\Omega^{5/6}\right)^2+1.2\sigma_I^2\Omega^{5/6}}} \right]}\textrm{,}
\\
&\left< {\Delta \theta_{1/2}}^2 \right>= \ln{\left[1+{1.2\sigma_I^2\Omega^{5/6}}/{\left(1+2.96\sigma_I^2\Omega^{5/6}\right)^{2}} \right]}\textrm{,}
\\
&\left< \Delta \theta_1 \Delta \theta_2 \right>= \ln{\left[1-{0.8\sigma_I^2\Omega^{5/6}}/{\left(1+2.96\sigma_I^2\Omega^{5/6}\right)^{2}} \right]}\textrm{,}
\\
&\left< \Delta x^2 \right>=\left< \Delta y^2 \right>= 0.33W_0^2\sigma_I^2\Omega^{-7/6}\textrm{,}
\end{aligned}
\end{gather}
where  $\Omega=\frac{kW_0^2}{2L}$, with $k$ being the optical wavenumber, $L$ the propagation distance, and $\sigma_I^2$ the scintillation index \cite{andrews2000scintillation}. The  scintillation index can be written
\begin{equation}
\sigma_I^2 = \exp{\left[ \frac{0.49\sigma_R^2}{\left(1+\zeta\sigma_R^{12/5}\right)^{7/6}} +  \frac{0.51\sigma_R^2}{\left(1+0.69\sigma_R^{12/5}\right)^{5/6}} \right]}-1\textrm{,}
\end{equation}
where the parameter $\zeta$ is set as 0.56 and 1.11 for uplink and downlink channels, respectively. In the above equation, $\sigma_R^2$ is the Rytov variance \cite{andrews2005laser},
\begin{equation}
\sigma_R^2 = 2.25k^{7/6}\int_{h_0}^{h_0+L} C_n^2\left(h\right)\left( h-h_0 \right)^{5/6}\,dh
\label{sigmaR}
\end{equation}
with $h_0$ the altitude above sea level of the ground station  and $C_n^2(h)$ the refraction index structure constant. Using the Hufnagel-Valley model, $C_n^2(h)$ is described by \cite{beland1993propagation}
\begin{align}
C_n^2(h)=&0.00594(v/27)^2(h\times10^{-5})^{10}e^{-\frac{h}{1000}}\nonumber\\
&+2.7\times 10^{-16}e^{-\frac{h}{1500}}+Ae^{-\frac{h}{100}}\textrm{,}
\label{lasteq}
\end{align}
where $v$ is the r.m.s. wind speed in ${m/s}$ and $A$ is the nominal value of $C_n^2(0)$ at sea level in ${m}^{-2/3}$.

Note, in the downlink satellite channel the beam-wandering effect is minor. This is largely because the beam-width on entry into the atmosphere from space is generally broader than the scale of the turbulent eddies \cite{andrews2005laser}. As such, with well-engineered designs the mean channel attenuation in the uplink and that in the downlink, can be anticipated to be  20-30dB and 5-10dB, respectively.

\section{CV-QKD with PSS and PAS}\label{PSSandPAS}

\subsection{Photon Addition and Photon Subtraction}
Before detailing our QKD protocol, we first discuss the experimental production of a multiple-Photon-Subtracted State (PSS) and a multiple-Photon-Added State (PAS).

A TMSV\footnote{A TMSV state can be created by applying the two-mode squeezing operator $S(r)$ to two vacuum states, where $r$ is the squeezing factor that satisfies $\sinh^2 r=\alpha^2$. The squeezing degree (in dB) of a TMSV state is given by $r_{[dB]} = -10\log_{10}{[\exp(-2r)]}$.} state  with mode $A$ and mode $B_0$ can be represented in the number (Fock) basis as
\begin{equation}
\left|\psi\right\rangle_{\rm{TMSV}}= \sum\limits_{n = 0}^\infty  {{a _n}} {\left| {n,n} \right\rangle _{A{B_0}}}
\end{equation}
with
\begin{equation}
{a _n} = \sqrt {\frac{{{\alpha ^{2n}}}}{{{{\left( {1 + {\alpha ^2}} \right)}^{n + 1}}}}}\ \textrm{,}
\end{equation}
where $\alpha^2$ is the mean photon number of the TMSV state. A TMSV state has the covariance matrix (CM)
\begin{equation}
\gamma_{\rm{TMSV}} = \left[ {\begin{array}{*{20}{cc}}
	\left(1+2\alpha^2\right)\bf{I} &2\sqrt{\alpha^4+\alpha^2}\sigma_z \\
	2\sqrt{\alpha^4+\alpha^2}\sigma_z&\left(1+2\alpha^2\right)\bf{I}  \\
	\end{array}}\right]\textrm{,}
\end{equation}
where ${\bf{I}}={\rm diag}(1,1)$, $\sigma_z={\rm diag}(1,-1)$, and we assume $\hbar=2$ so that the vacuum noise is 1.

A PSS can be generated by inserting a mode, say $B_0$, of state $\left|\psi\right\rangle_{\rm{TMSV}}$ and an ancillary mode $C_0=\left|0 \right\rangle$ into the two inputs of a beam splitter (see red box of Fig.~\ref{fig:qkd}a). Denoting  the transmissivity of the beam splitter as $T_S$, and its two output modes   as $B$ and $C$, a PSS with $N$ photons subtracted is produced when  $C=\left|N \right\rangle$. The PSS has the form
\begin{equation}
{\left| \psi  \right\rangle _{\rm{PSS}}} = \frac{(-1)^N}{{\sqrt {{P_{S,N}}} }}\sum\limits_{n = N}^\infty  {{a_n}r_{n,N}^{T_S}} {\left| {n,n - N} \right\rangle _{AB}}\ \textrm{,}
\label{substate}
\end{equation}
where
\begin{equation}
r_{n,N}^T = \sqrt {\left( {\begin{array}{*{20}{c}}
		n  \\
		N  \\
		\end{array}} \right)} {(\sqrt T) ^{n - N}}{\sqrt {1 - T} ^N}\textrm{.}
\end{equation}
 The probability for this outcome can be determined as
\begin{equation}
P_{S,N} = \frac{\left[ \alpha^2\left(1 - T_{S} \right)\right] ^N }{\left( 1+\alpha^2-\alpha^2T_{S}\right)^{N+1} }\ .
\label{subProb}
\end{equation}
Such states possess a CM given by
\begin{align}
\gamma_{S} &= \left[ {\begin{array}{*{20}{cc}}
	x\bf{I} &z\sigma_z \\
	z\sigma_z&y\bf{I}  \\
	\end{array}} \right]\nonumber\\
&=
\left[ {\begin{array}{*{20}{cc}}
	\left( 1+2\frac{N+T}{1-T}\right)\bf{I} &\left( 2\sqrt{T}\frac{N+1}{1-T}\right)\sigma_z \\
	\left( 2\sqrt{T}\frac{N+1}{1-T}\right)\sigma_z &\left( 1+2\frac{(N+1)T}{1-T}\right)\bf{I} \\
	\end{array}} \right]\textrm{,}
\label{CMS}
\end{align}
where $T=\frac{\alpha^2}{1+\alpha^2}T_S$.

A PAS can be produced by replacing the aforementioned state $C_0$ with $\left|N \right\rangle$  (see red box of Fig.~\ref{fig:qkd}b). It has the form
\begin{gather}
\begin{aligned}
{\left| \psi  \right\rangle_{\rm{PAS}}} &=  \frac{(-1)^N}{{\sqrt {{P_{A,N}}} }}\sum\limits_{n = 0}^\infty  {{a _n}r_{n+N,N}^{T_S}} {\left| {n,n + N} \right\rangle _{AB}}\\
&=   \frac{(-1)^N\sqrt{(1+\alpha^{-2})^N}}{{\sqrt {{P_{A,N}}} }}\sum\limits_{n = N}^\infty  {{a_n}r_{n,N}^{T_S}} {\left| {n-N,n} \right\rangle _{AB}}\ \textrm{,}
\label{addstate}
\end{aligned}
\end{gather}
and is created when $C=\left|0 \right\rangle$. This outcome has a probability
\begin{equation}
P_{A,N} = \frac{\left[ (\alpha^2+1)\left(1 - T_{S} \right)\right] ^N }{\left( 1+\alpha^2-\alpha^2T_{S}\right)^{N+1} }.
\label{addProb}
\end{equation}

A PAS possesses a CM similar  to $\gamma_{S}$ except for the values of the variances of the two modes being swapped., i.e.,
\begin{equation}
\gamma_{A} = \left[ {\begin{array}{*{20}{cc}}
	y\bf{I} &z\sigma_z \\
	z\sigma_z&x\bf{I}  \\
	\end{array}}\right].
\end{equation}

For simplicity we assume that both $\left|N \right\rangle$ and $\left|0 \right\rangle$ are produced with unit efficiency. In this case, the overall successful probability for the photon subtraction and that for the photon addition are $P_{S,N}$ and $P_{A,N}$, respectively.

\subsection{CV-QKD Protocol and Secret Key Rate}
\begin{figure}
	\includegraphics[width=.48\textwidth]{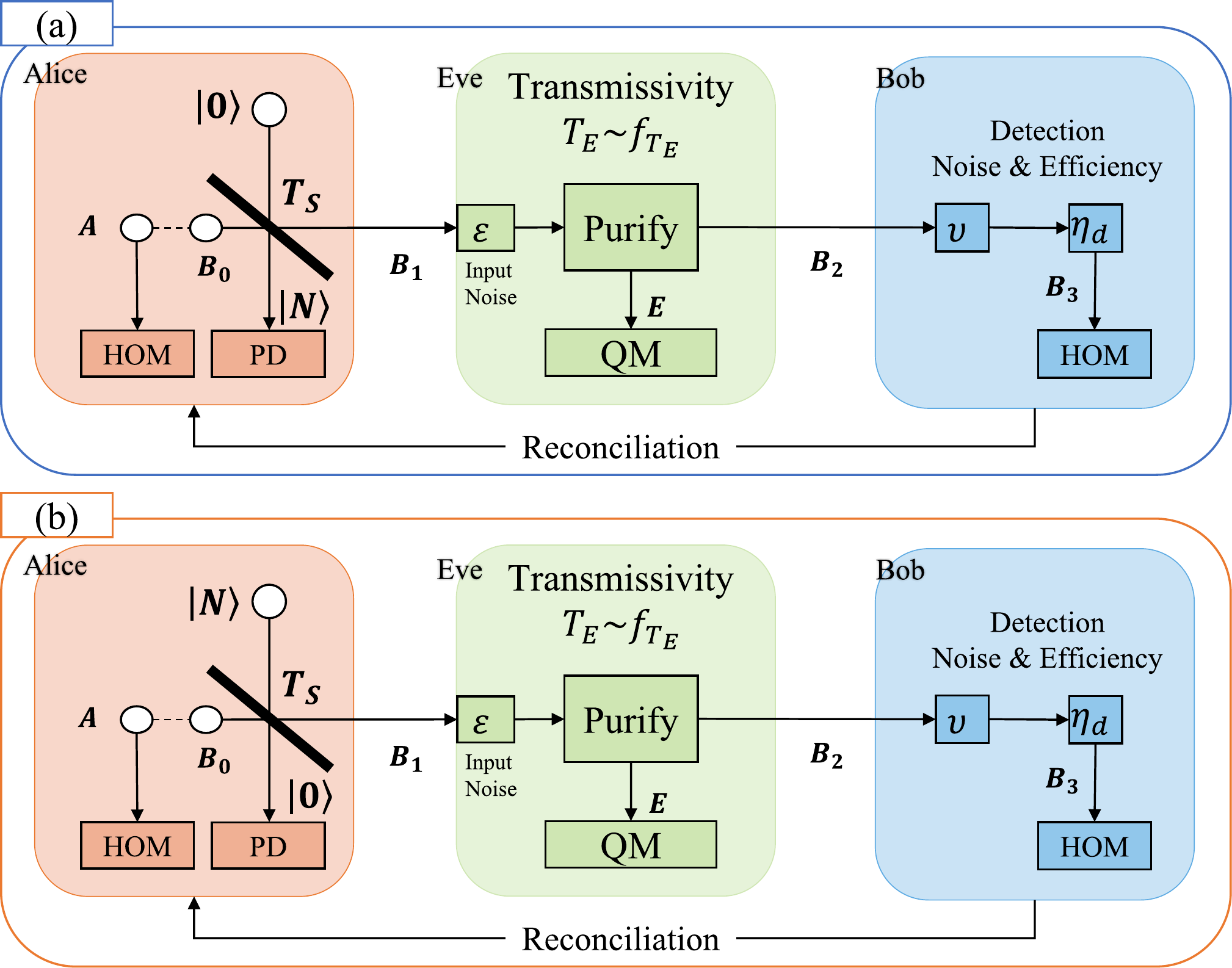}
	\centering
	\caption{\textbf{(a)} The T-PS scheme. \textbf{(b)} The T-PA scheme. (QM: quantum memory. HOM: homodyne detector. PD: photon detector.)}
	\label{fig:qkd}
\end{figure}

In this work we consider an entanglement-based CV-QKD protocol (homodyne at Alice and Bob) with reverse reconciliation. We study only non-Gaussian operations at the transmitter side.\footnote{We focus on the transmitter side because we consider the use of photonic engineering coupled to \emph{future} quantum memory at the transmitter will ultimately lead to better real-time outcomes relative to photonic-engineering only at the receiver side. For example, non-Gaussian states stored in the transmitter quantum memory can be used  `on-demand' in future QKD set-ups.}
We consider two schemes, namely the T-PS and the T-PA schemes.
As illustrated in Fig.~\ref{fig:qkd}, in both schemes Alice first prepares her TMSV state $\rho_{AB_0}$. She then produces the PSS (in T-PS scheme) or the PAS (in T-PA scheme) by performing corresponding non-Gaussian operation to $\rho_{AB_0}$.
The resultant state of the non-Gaussian operation is labeled with $\rho_{AB_1}$. Alice will  send mode $B_1$ to Bob.

Note that, comparing Eq.~(\ref{subProb}) and (\ref{addProb}) we find that $P_{A,N}$ is larger than $P_{S,N}$ by a factor of $\left( 1+\alpha^{-2}\right)^N$.
We also note from Eq.~(\ref{substate}) and Eq.~(\ref{addstate}) that for a TMSV state, subtracting photons from one mode yields the same state as adding photons to another mode (up to a normalization factor) \cite{navarrete2012enhancing}. In fact, this latter method of `photon subtraction' retains the higher probability given by Eq.~(\ref{addProb}). Considering this advantage we assume that in our T-PS scheme the PSS is created by adding photons to mode $A$ of the TMSV state $\rho_{AB_0}$. The resultant PSS has the CM, $\gamma_{S}$, given by Eq.~(\ref{CMS}).
Under the assumptions made here, the probabilities for T-PS and its corresponding T-PA are the same. The reader should scale the calculations to follow, if they wish to account for some non-equal probabilities (e.g. a non-unity efficiency of producing the state $\left|N \right\rangle$).


A frequently used attack is the entanglement cloning attack \cite{weedbrook2012continuous,laudenbach2018continuous,grosshans2003virtual}. Under this attack it is usually assumed that Eve can obtain a purification of Alice and Bob's system, so that she can attain the maximal quantum information. However, for non-Gaussian states the premise of this attack is not, in general, compatible with the Gaussification assumption under which Eve replaces the non-Gaussian state $\rho$ with a Gaussian state $\rho^G$ that has the same CM, since one can show that $\rho^G$ is not a pure state (i.e. the entanglement cloning attack is not optimal for non-pure state). Therefore, we do not consider here any physical implementation of Eve's attack: we simply assume  that Eve Gaussifies the state $\rho_{AB_1}$ and purifies the Gaussified state $\rho_{AB_2}$ with her ancillary system $\rho_{E}$.

Let us denote the channel transmissivity as $T_E$ and the input excess noise as $\varepsilon$. We assume Eve hides herself by mimicking the anticipated noise conditions  for a given realization of $T_E$, based on Eqs. (\ref{Cheq1}) to (\ref{lasteq}). Eve will interact her ancillary mode with every incoming mode, she will then store her  mode in a quantum memory, performing a joint measurement on her  system after the reverse reconciliation process.

After receiving a mode, Bob will perform an imperfect homodyne detection with a detection thermal noise $\nu$ and an efficiency $\eta_d$. This imperfection can be modeled by first interacting mode $B_2$ with one mode of a TMSV state ($G_0-H$) with variance $\nu$ at a beam splitter with transmissivity $\eta_d$ \cite{hosseinidehaj2016cv}. Denoting $G$ and $B_3$ as the output modes of such a beam splitter, mode $B_3$ will then be injected into a perfect homodyne detector. Under this scenario we assume  Eve does not have access to Bob's detection device.

In the asymptotic limit for key size, the secret key rate for our non-Gaussian state protocol is lower bounded by \cite{navascues2006optimality}
\begin{equation}
K\geq P_{N}\left[ \eta_rI(A\negmedspace:\!B_3) - \chi(E\negmedspace:\!B_3)\right]\textrm{,}
\label{keyrate}
\end{equation}
where $P_{N}=P_{A,N}$ is the successful probability for adding $N$ photons to the TMSV state, $\eta_r$ is the reverse reconciliation efficiency, $I(A\negmedspace:\!B_3)$ is the classical mutual information  between Alice and Bob, and $\chi(E\negmedspace:\!B_3)$ is the Holevo bound for Eve's information (which is the attainable quantum information given Bob's measurements).

Alice and Bob's mutual information is calculated by
\begin{align}
I(A\negmedspace:\!B_3)&=H(A)-H(A|B_3)\nonumber\\
&=\frac{1}{2}\log_2{\frac{V_A}{V_{A|B_3}}}\textrm{,}
\label{AandB}
\end{align}
where $V_{A}$ is the variance of Alice's mode, and $V_{A|B_3}$ is the variance of Alice's mode conditioned on Bob's homodyne measurement.
For the Holevo bound, since Eve holds a purification of $\rho_{AB_2}$, and after Bob's measurement the system $\rho_{AEGH}$ is pure, we have
\begin{align}
\chi(E\negmedspace:\!B_3) &= S(E)-S(E|B_3)\nonumber\\
&=S(AB_2)-S(AGH|B_3)\nonumber\\
&=\sum_{i=1}^{2} g\left( \lambda_i\right) - \sum_{j=3}^{5} g\left( \lambda_j\right)\textrm{,}
\label{EandB}
\end{align}
where
$
g(x)=\frac{x+1}{2}\log_2\frac{x+1}{2}-\frac{x-1}{2}\log_2\frac{x-1}{2}
$, and $\lambda_i$'s and $\lambda_j$'s are the symplectic eigenvalues of the CM characterizing the state $\rho_{AB_2}$, and the CM characterizing the state $\rho_{AGH}$ conditioned on Bob's measurement, respectively.

\subsection{Evolution of the Covariance Matrix}
For calculation of the secret key rate we need to calculate $I(A\negmedspace:\!B_3)$ and $\chi(E\negmedspace:\!B_3)$, which are determined by the CMs of the states $\rho_{AB_2}$ and $\rho_{AGHB_3}$. Noticing the evolution of the CM follows the same procedure for both T-PS and T-PA schemes,  we use the T-PS scheme as an example to derive the CMs.

The PSS $\rho_{AB_1}$ prepared by Alice has the CM $\gamma_{AB_1}$ given by Eq.~(\ref{CMS}). After the channel, the state $\rho_{AB_1}$  evolves to  $\rho_{AB_2}$, the CM of which is
\begin{align}
\gamma_{AB_2} &= \left[ {\begin{array}{*{20}{cc}}
	x\bf{I} &\sqrt{T_E}z\sigma_z \\
	\sqrt{T_E}z\sigma_z&\left[T_E(y+\varepsilon)+(1-T_E)\right]\bf{I}  \\
	\end{array}}\right]\nonumber\\
&=\left[ {\begin{array}{*{20}{cc}}
	x\bf{I} &z'\sigma_z \\
	z'\sigma_z&y'\bf{I}  \\
	\end{array}}\right].
\end{align}
The symplectic eigenvalues of the above CM are
\begin{equation}
\lambda_{1,2}=\frac{1}{2}\left[ \sqrt{(x+y')^2-4z'^2}\pm(y'-x)\right].
\end{equation}
The CM of the state before Bob's detection ($\rho_{AGHB_3}$) is
\begin{align}
\gamma_{AGHB_3} &= \left[ {\begin{array}{*{20}{cc}}
	\gamma_{AGH} &\sigma_{AGHB_3} \\
	\sigma_{AGHB_3}^T&\gamma_{B_3}\\
	\end{array}}\right]\textrm{,}
\end{align}
where
\begin{gather}
\gamma_{AGH} = \left[ {\begin{array}{*{20}{cc}}
	x\bf{I}&C_{AG}\sigma_z&0\\
	C_{AG}\sigma_z&\left[\eta_d \nu+(1-\eta_d)y'\right]\textbf{I}&C_{GH}\sigma_z\\
	0&C_{GH}\sigma_z&\nu \bf{I}
	\end{array}}\right]\textrm{,}\\
\sigma_{AGHB_3}=\left[\begin{array}{*{20}{cc}}
\sqrt{\eta_d}z'\sigma_z\\
\sqrt{(1-\eta_d)\eta_d}(\nu-y')\sigma_z\\
\sqrt{(1-\eta_d)(\nu^2-1)} \sigma_z
\end{array}\right]\textrm{,}
\end{gather}
$C_{AG}=-\sqrt{1-\eta_d}z'$, $C_{GH}=\sqrt{\eta_d(\nu^2-1)}$, and \begin{equation}
\gamma_{B_3}=\left[\eta_d y'+(1-\eta_d)\nu\right]\textbf{I}=y''\textbf{I}.
\end{equation}
Suppose Bob homodynes the $x$-quadrature of mode $B_3$, the CM of the state $\rho_{AGH}$ conditioned on such measurement is
\begin{equation}
\gamma_{AGH|B_3}=\gamma_{AGH}-\frac{1}{y''}\sigma_{AGHB_3}X\sigma_{AGHB_3}^T\textrm{,}
\end{equation}
where $X=\textrm{diag}(1,0)$. Let $\lambda_{3}$, $\lambda_{4}$, and $\lambda_{5}$ be the symplectic eigenvalues of the CM $\gamma_{AGH|B_3}$;  
then the Holevo bound for Eve's information $\chi(E\negmedspace:\!B_3)$ can be calculated by putting the symplectic eigenvalues ($\lambda_1$ to $\lambda_5$) into Eq.~(\ref{EandB}).

The mutual information between Alice and Bob in Eq.~(\ref{AandB}) can now be expressed as
\begin{equation}
I(A\negmedspace:\!B_3)=\frac{1}{2}\log_2{\frac{1}{1-\frac{\eta_d T_E  z^2}{\eta_d T_E  xy+cx}}}\ \textrm{,}
\label{AandBre}
\end{equation}
where $c=\eta_d\left[(1-T_E)+T_E \varepsilon \right] + (1-\eta_d)\nu$.

We are ready to use Eq.~(\ref{keyrate}) to calculate the secret key rate for a given channel transmissivity $T_E$.
Before doing this let us note that if we denote
$f_{T_E}$ to be the PDF of $T_E$,  the average secret key rate over the Earth-satellite channel is
\begin{equation}
\overline K=\int_{0}^{T_0}f_{T_E}(T)K(T)\,dT\textrm{,}
\label{avgkeyrate}
\end{equation}
where  $K(T)$  is the secret key rate for a transmissivity $T$.

\section{Simulation Results}\label{results}
\begin{figure}[t]
	\includegraphics[width=0.37\textwidth]{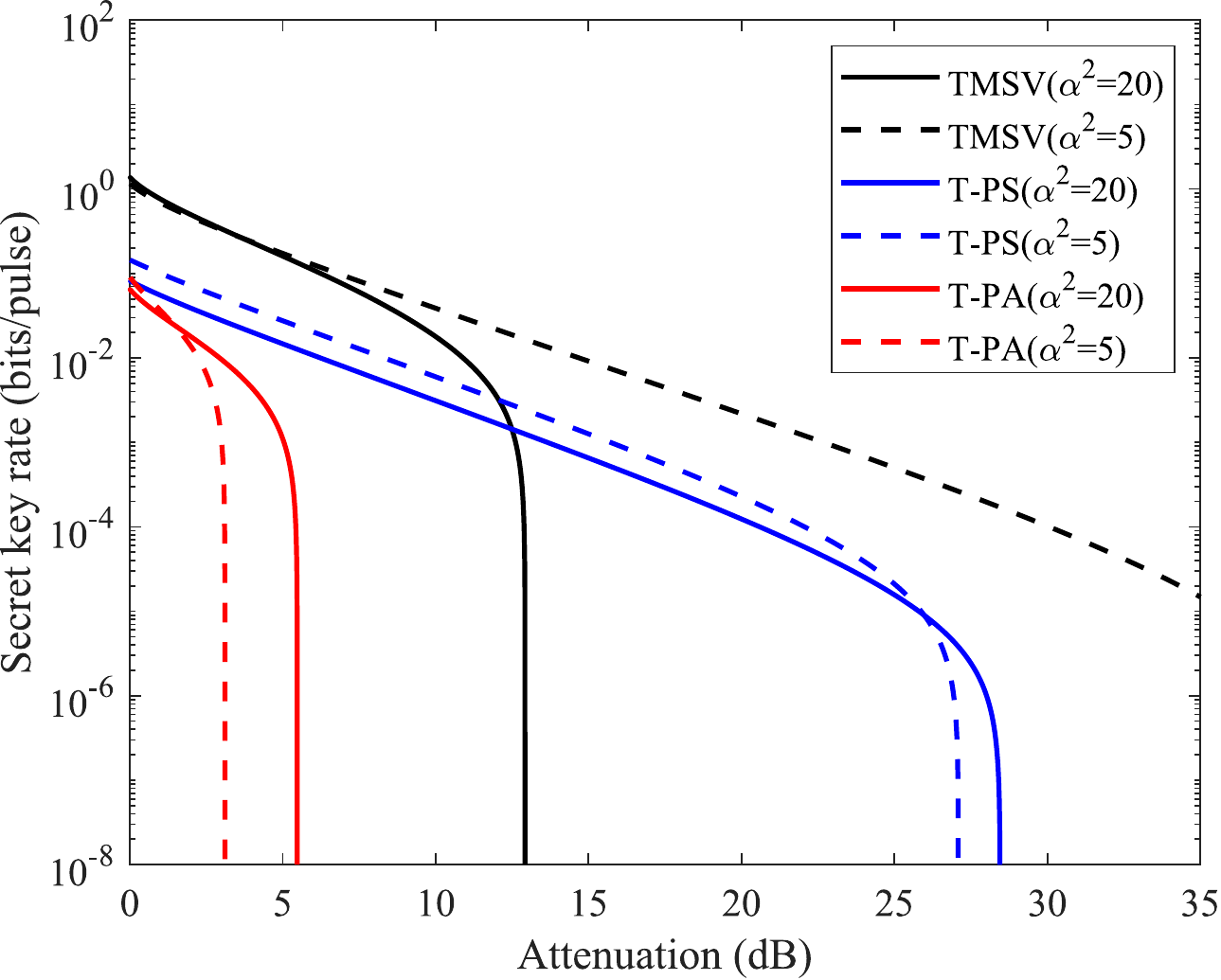}
	\centering
	\caption{Secret key rate over the fixed-attenuation channel with $\alpha^2 = 20$ (solid) and $\alpha^2 = 5$ (dashed). Here we set $T_S = 0.7$ and $N=1$. Note that for the non-TMSV schemes, the $\alpha^2$ refers to the TMSV state that was initially utilized in the scheme.}
	\label{FixedRate}
\end{figure}
\begin{figure}[ht]
	\includegraphics[width=0.37\textwidth]{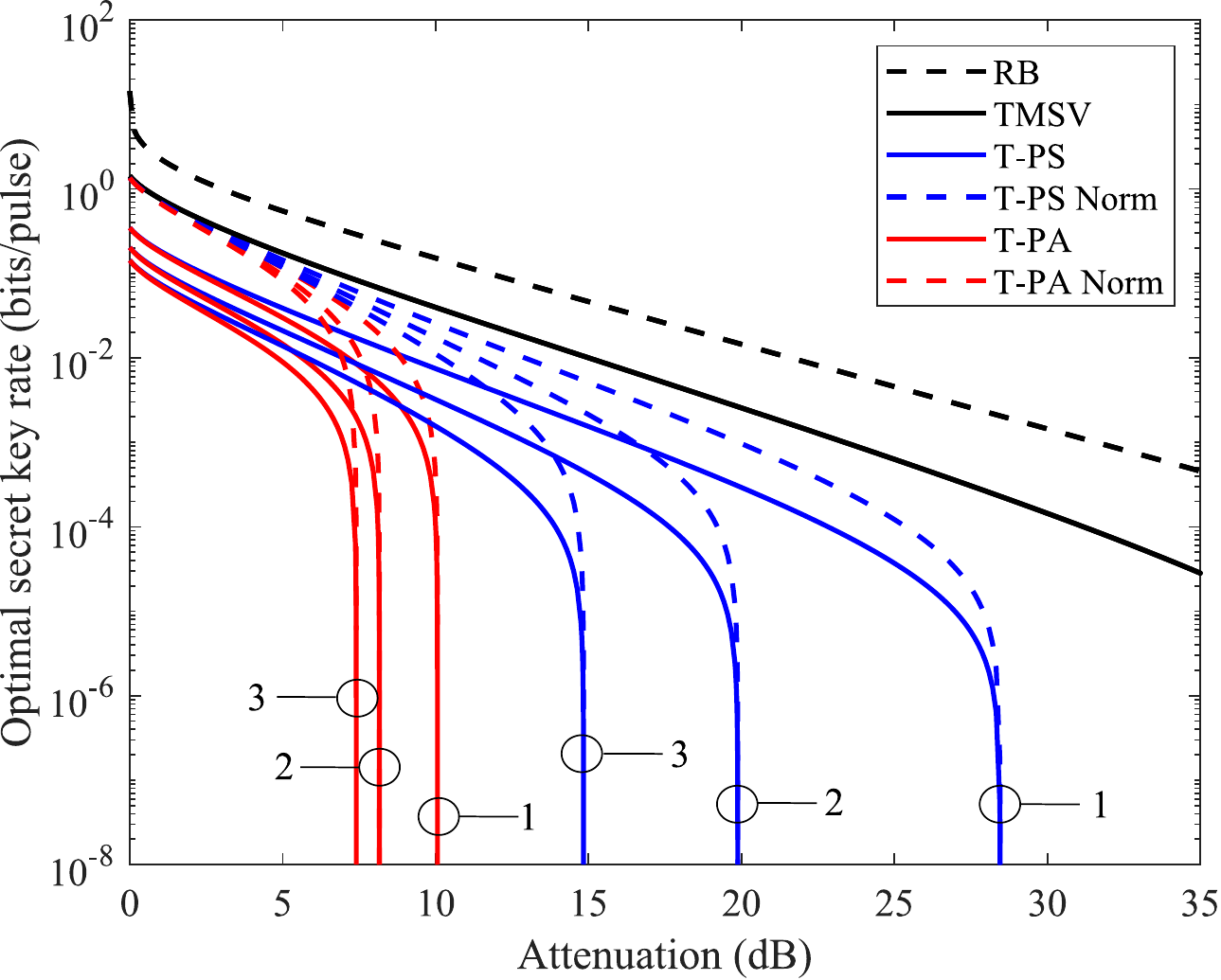}
	\centering
	\caption{Optimal secret key rate over the fixed-attenuation channel. The RB is given by $C=-\log_2(1-T_E)$. The blue and red dashed curves represents the scenario where a quantum memory device is at the transmitter, so that Alice can prepare her PSS or PAS in advance. In such a scenario we can assume the probability of creating a PSS or a PAS is one (the curves labeled `Norm' indicate such a scenario).}
	\label{OptRate}
\end{figure}
\begin{figure}
	\includegraphics[width=0.37\textwidth]{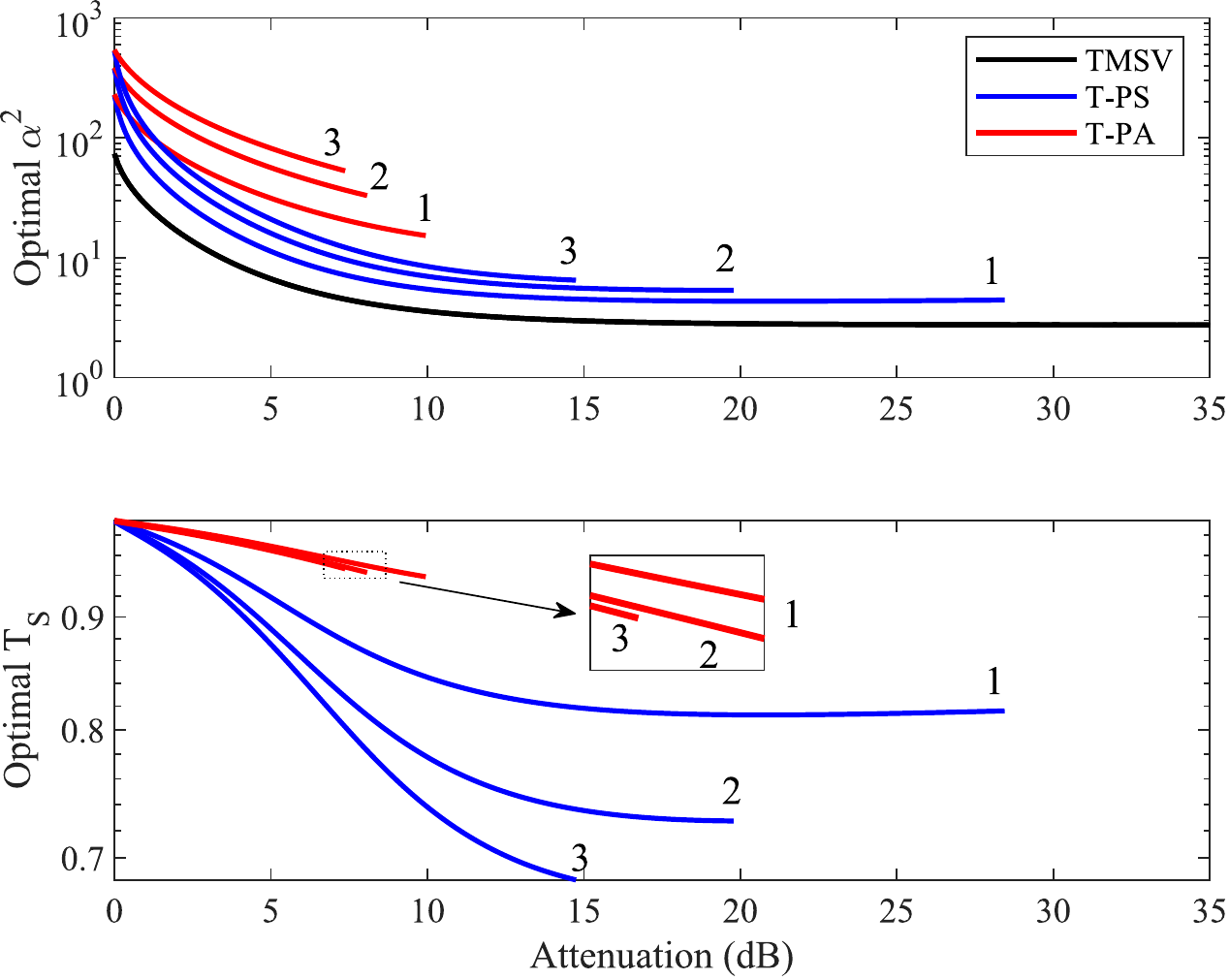}
	\centering
	\caption{Optimal $\alpha^2$ (top) and $T_S$ (bottom) vs. channel attenuation.}
	\label{OptPara}
\end{figure}

The noise parameters we adopt are $\varepsilon=0.1$ and $\nu=1.1$ (both in vacuum noise units), which are roughly $10\%$ higher than the experimental data of \cite{huang2013performance}. We also adopt the efficiency parameters from \cite{lim2017continuous}, namely, $\eta_r=0.95$ and $\eta_d=0.68$. Using the initial TMSV state (i.e.  no photonic engineering) as a reference, we first compare the performance of different schemes over fixed-attenuation channels. We consider two  different entanglement levels. We set the mean photon number of the TMSV state, $\alpha^2$, equal to 20 and 5, where the former
value represents a higher entanglement level. The transmissivity of the beam splitter for photon addition is set as $T_S = 0.7$.

As illustrated in Fig.~\ref{FixedRate}, at  higher entanglement (solid curves) the T-PS scheme has a higher key rate at high attenuation.
At lower entanglement (dashed curves), however, the TMSV scheme has the highest secret key rate over the \emph{entire} attenuation range. Comparing just the curves of the T-PS scheme at the different entanglement levels we can see that the lower entanglement curve has a higher secret key rate when the channel attenuation channel is less than 26dB. This means, that in practice the value of $\alpha^2$ for a given channel transmissivity $T_E$ can be set so as to optimize the secret key rate. A similar discussion applies to the T-PA scheme.
Such optimization can be implemented through some channel co-measurement (e.g a coherent beam sent simultaneously with the quantum signal) and a closed-feedback loop. Note, that we also can optimize the key rate via setting the value of $T_S$.

If indeed we optimize $\alpha^2$  and $T_S$ (simultaneously) along these lines for each scheme the results are different. Fig.~\ref{OptRate} shows the secret key rate with optimized parameters for the fixed-attenuation channel, and Fig.~\ref{OptPara} illustrates the corresponding optimal parameters as a function of attenuation. Note, in Fig.~\ref{OptRate}, and subsequent figures, the digits `1', `2' and `3' near each curve illustrate the number of photons ($N$) being subtracted (or added), and RB represents the `Repeaterless Bound', which is the upper bound for the channel capacity of a pure-loss channel without a repeater \cite{pirandola2017fundamental,wilde2017converse} \footnote {See \cite{pirandola2018theory} for further discussion on the history of this bound.}. In Fig.~\ref{OptRate} the TMSV scheme has the highest key rate across all attenuation values. Even in the situation where there is a quantum memory set-up at the transmitter, the T-PS and T-PA schemes still have lower key rates than the TMSV scheme. We also note that the for a given quantum memory scenario and a given $N$, the T-PA scheme always has worse performance than the T-PS scheme. This is mainly because that Eve can obtain more information from the PAS than the PSS.

We now move onto the investigation of the CV-QKD protocols over Earth-satellite fading channels. For the fading channel we adopt the parameters from the experiment in \cite{vasylyev2016atmospheric}, namely $W_0=20mm$, $r_0=40mm$, and $h_0=0m$; and set the wavelength of the beam to $809nm$. By varying $\sigma_I^2$ and $L$ we study the performance of our schemes in various loss conditions. These conditions are characterized with the mean channel transmissivity
\begin{equation}
\overline T_E=\int_{0}^{T_0}f_{T_E}(T)T\,dT.
\label{avgTe}
\end{equation}

Since there are no closed-form solutions for the PDF of the channel transmissivity $T_E$, we use a Monte Carlo algorithm to simulate the channel. In this case, Eq.~(\ref{avgkeyrate}) and Eq.~(\ref{avgTe}) are approximated by
\begin{equation}
\overline K\approx\frac{1}{N_s}\sum_{n}^{N_s} K(T_n)\textrm{,}
\label{avgkeyrateMC}
\end{equation}
and
\begin{equation}
\overline T_E\approx\frac{1}{N_s}\sum_{n}^{N_s} T_n\ \textrm{,}
\label{meanTE}\
\end{equation}
respectively, where $T_n$ is channel transmissivity sample generated by the Monte Carlo algorithm, and $N_s$ is the number of transmissivity samples. We set $N_s=2^{20}$ for each simulation run.

\begin{figure}[t]
	\includegraphics[width=0.37\textwidth]{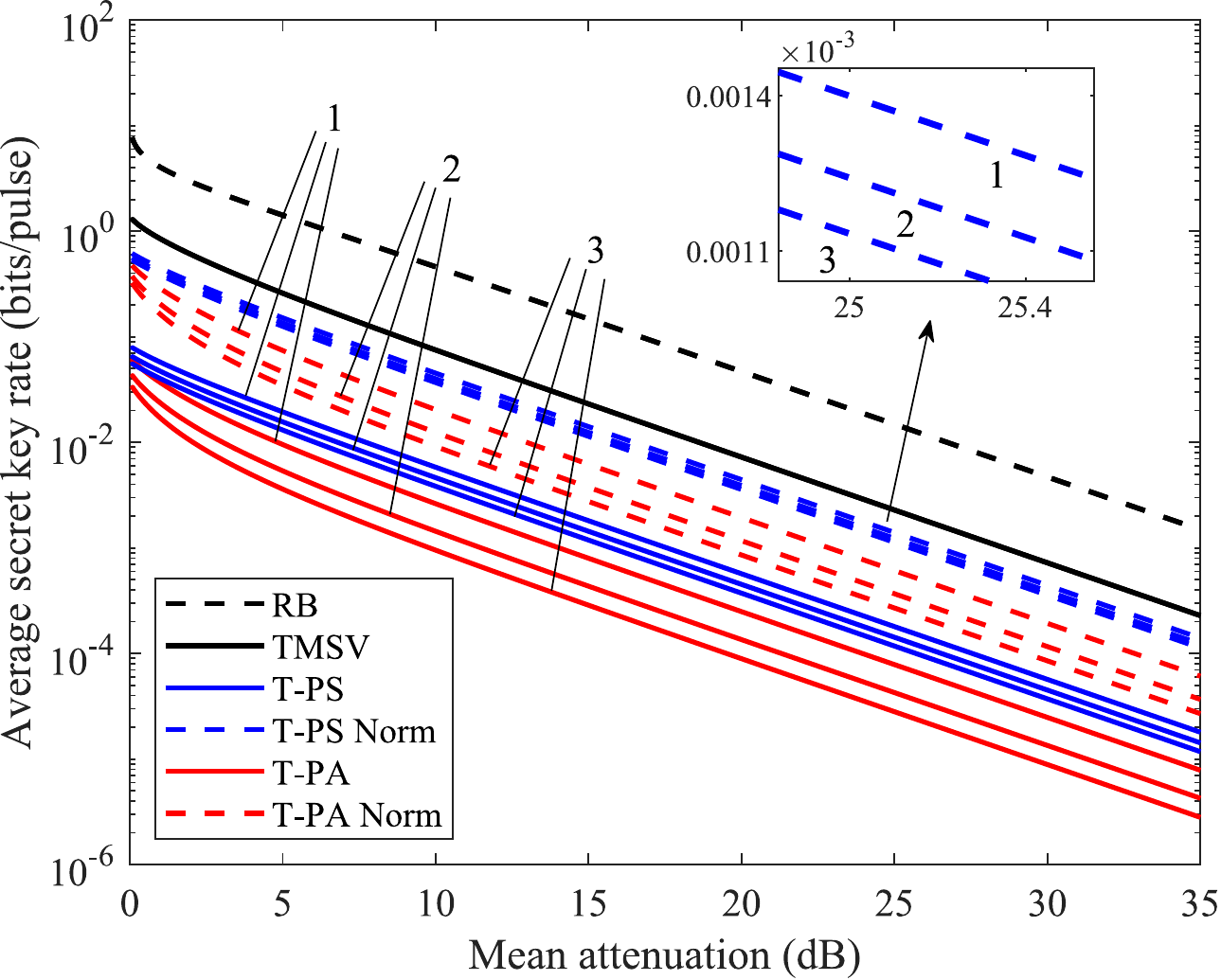}
	\centering
	\caption{Average secret key rate over the Earth-satellite channel with only the beam-wandering effect considered. Here the RB is averaged and given by $C= -\int_{0}^{T_0} f_{T_E}(T)\log_2(1-T)\,dT$. ($\alpha^2 = 20$ and $T_S = 0.7$)}
	\label{AvgRateBmwd}
\end{figure}
\begin{figure}
	\includegraphics[width=0.37\textwidth]{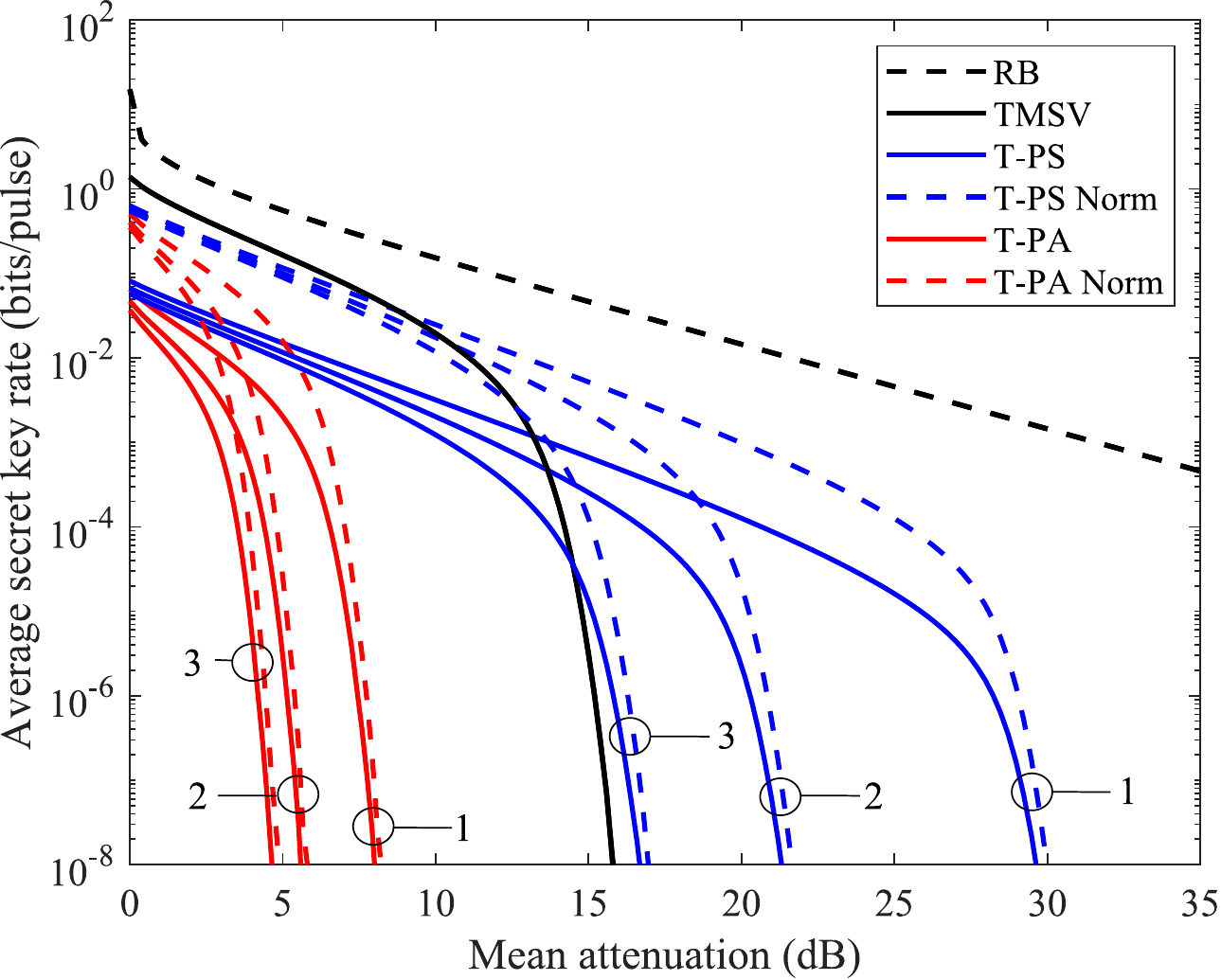}
	\centering
	\caption{Average secret key rate over the Earth-satellite channel with all the three effects considered. Here the RB is approximated by $C= -\sum_{n=0}^{N_s} \log_2(1-T_n)/N_s$. ($\alpha^2 = 20$ and $T_S = 0.7$)}
	\label{AvgRateFixed}
\end{figure}
\begin{figure}
	\includegraphics[width=0.37\textwidth]{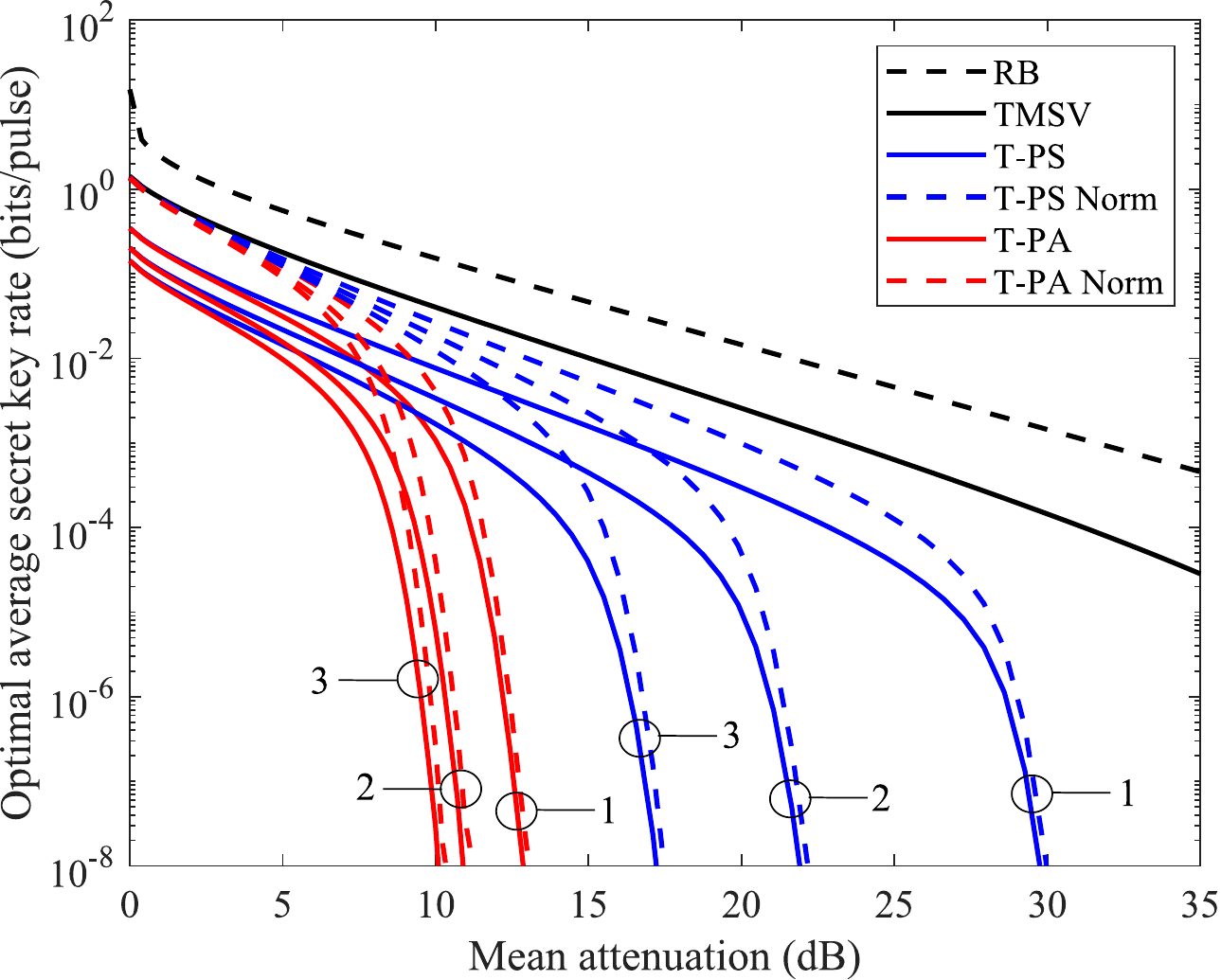}
	\centering
	\caption{Average key rate optimized for mean channel transmissivity. We consider the Earth-satellite channel with all the three effects. Here the RB is the same as Fig.~\ref{AvgRateFixed}.}
	\label{AvgRate}
\end{figure}
\begin{figure}
	\includegraphics[width=0.37\textwidth]{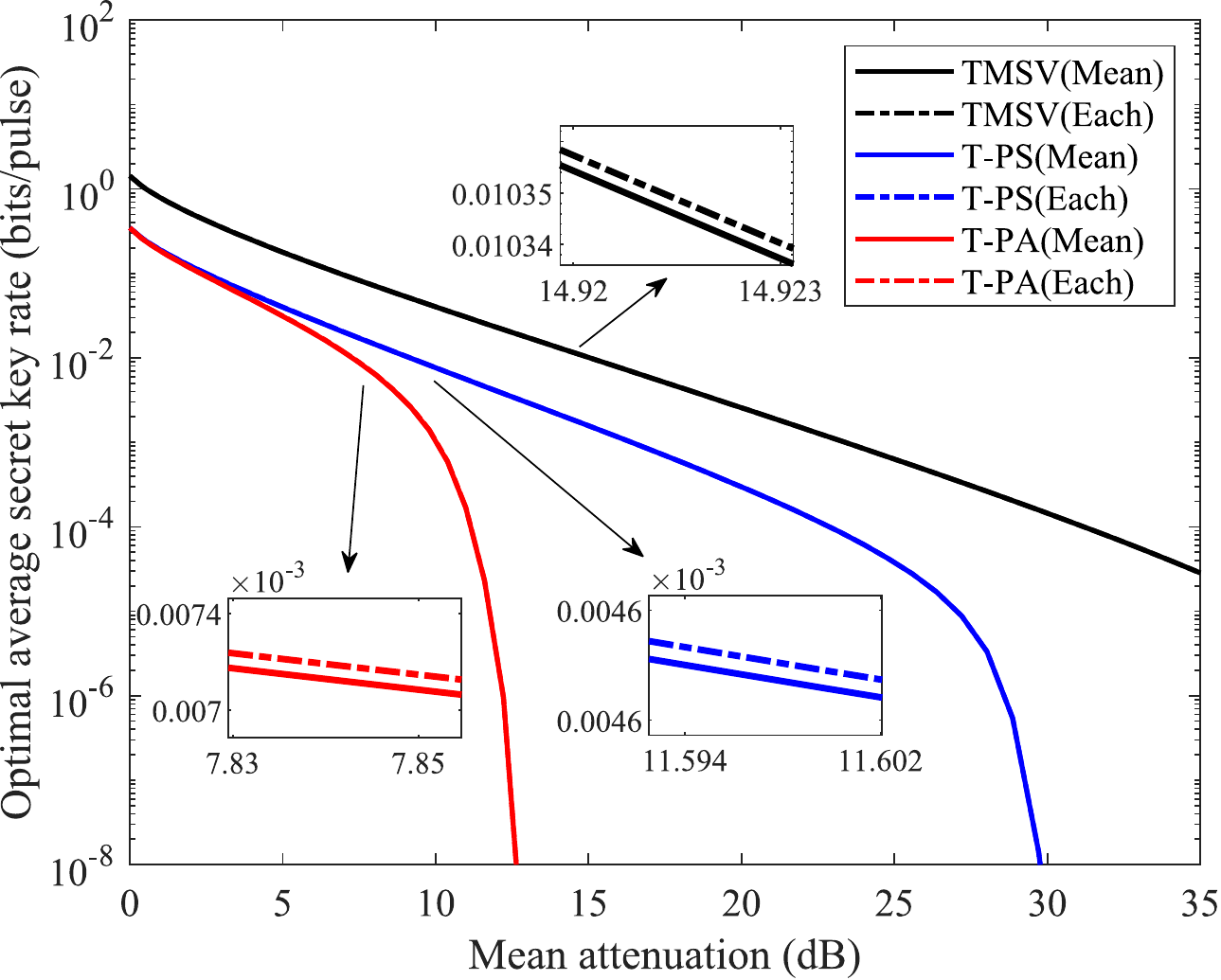}
	\centering
	\caption{Average secret key rate optimized for \emph{each} measured channel transmissivity (dashed-dot curves), and optimized for \emph{mean} channel transmissivity (solid curves). Here we consider all the three channel effects. For conciseness only the scenario that $N=1$ is illustrated.}
	\label{AvgRateEach}
\end{figure}

We first study the impact of the updated channel model. To do so, we consider the simplest case of no channel estimation or optimization. Fig.~\ref{AvgRateBmwd} illustrates the scenario where we only consider the beam-wandering effect over the Earth-satellite channel (in this scenario the closed-form solution to $f_{T_E}$ is tractable). Here we assume $\frac{r_0}{W_1}=\frac{r_0}{W_2}=2$. One can see the average secret key rate of all the schemes drops gently with increasing mean attenuation. When we assume the atmospheric turbulence does not change the size or shape of the beam-profile, the maximal attainable transmissivity $T_0$ would be fixed for a given propagation distance. Consequently, it is always possible to sample (albeit with low probability) a large transmissivity regardless of the mean attenuation value.

Taking the beam-broadening and the beam-deformation effects caused by the turbulence into consideration, in Fig.~\ref{AvgRateFixed} the average key rate is plotted against mean channel attenuation. These two effects bring drastic changes. Assuming the size and shape of the beam-profile are distorted by the  turbulence, a random mismatch would occur between the beam-profile and the detector aperture. In this case, the maximal attainable transmissivity $T_0$ itself will follow a certain probability distribution, making the probability of having a large transmissivity significantly smaller than the previous scenario.

For parameter optimization we consider two schemes. In our first scheme we estimate the mean transmissivity $\overline T_E$ of the channel, and then optimize the parameters $\alpha^2$ and $T_S$ based on the anticipated $\overline T_E$. The results are illustrated in Fig.~\ref{AvgRate}. Here for a given $\overline T_E$ we directly use the results from Fig.~\ref{OptPara} for the optimal parameter settings. Comparing Fig.~\ref{OptRate} and Fig.~\ref{AvgRate} one can see that all the schemes have better attenuation tolerance over the Earth-satellite channel. This is  due to the fading property of the Earth-satellite channel. Simply put, the probability of having a transmissivity larger than its mean value is always non-zero.

We next consider a more complex system, where we measure the channel transmissivity within each coherence time window and optimize the parameters based on this measurement. In this case, the parameters are optimized for each $T_n$. The two optimization schemes are compared in Fig.~\ref{AvgRateEach}. The measurement-based scheme (for each $T_n$) offers only a slight improvement to the secret key rate. Considering the complexity of such a scheme, the mean-value based optimization scheme would be a good practical choice for physical implementation.

An important result is that for satellite communications the single-PSS has the highest average key rate amongst the non-Gaussian states we have considered, making it our preferred candidate for non-Gaussian state based CV-QKD protocols.
Focusing on the single-PSS, over the downlink channel (5-10dB attenuation) a mean photon number $\alpha^2\sim10$ of the initial TMSV state will suffice to achieve a key rate of $10^{-2}$ bits/pulse. Such a mean photon number is equivalent to a 16dB squeezed TMSV state, which is experimentally achievable at the moment. Over the uplink channel (20-30dB attenuation) the optimal $\alpha^2$ is between 5 to 10. The corresponding optimal secret key rate is around $10^{-4}$ bits/pulse. One should also notice that for non-Gaussian states we employed the lower bound to evaluate their performance, so the actual secret key rate will likely be higher.

\section{Conclusion}
We have studied the use of PSSs and PASs in the context of CV-QKD - investigating the lower-bounds on optimal secret key rates delivered by such states when the photonic engineering  is performed at the transmitter.
Our results highlight the role played by deployment scenarios  in determining the optimal photonic engineered state, as well as highlighting the role  atmospheric turbulence has on the predicted  key rates.
Beyond the impact on the secure key rates,
the results shown here could also be important for future space missions in which non-Gaussian states will be used for other applications.

\section{Acknowledgement}
The authors acknowledge financial support from Northrop Grumman Corporation. Mingjian He is partially supported by a Postgraduate award through the China Scholarship Council. The authors thank N. Hosseinidehaj and the anonymous referees for their comments on an earlier version of this manuscript.


\end{document}